\documentclass[aps,prl,twocolumn,showpacs,groupedaddress]{revtex4}  
\usepackage{graphicx}  
\usepackage{dcolumn}   
\usepackage{bm}        
\usepackage{amssymb}   

\def\lsim{\mathrel{\rlap{\lower4pt\hbox{$\sim$}}
    \raise1pt\hbox{$<$}}}                

\newcommand{\ppb}{\mbox{\ensuremath{p\bar p}}}
\newcommand{\qqb}{\mbox{\ensuremath{q\bar q}}}

\newcommand{\invpb}{pb$^{-1}$}
\newcommand{\Gev}{GeV}

\newcommand{\met}{\mbox{\ensuremath{\slash\kern-.7emE_{T}}}}
\newcommand{\mht}{\mbox{\ensuremath{\slash\kern-.7emH_{T}}}}

\newcommand{\pt}{\mbox{\ensuremath{p_{T}}}}
\newcommand{\etadet}{\mbox{\ensuremath{\vert\eta_\mathrm{det}\vert}}}

\newcommand{\dphimin}{\mbox{\ensuremath{\Delta\Phi_\mathrm{min}}}}
\newcommand{\dphimax}{\mbox{\ensuremath{\Delta\Phi_\mathrm{max}}}}

\begin{document}


\hspace{5.2in} \mbox{Fermilab-Pub-06/233-E}

\title{Search for scalar leptoquarks in the acoplanar jet 
topology\\ in \ppb\ collisions at $\sqrt{s}=1.96$\,TeV}

%
\author{                                                                      
V.M.~Abazov,$^{36}$                                                           
B.~Abbott,$^{76}$                                                             
M.~Abolins,$^{66}$                                                            
B.S.~Acharya,$^{29}$                                                          
M.~Adams,$^{52}$                                                              
T.~Adams,$^{50}$                                                              
M.~Agelou,$^{18}$                                                             
J.-L.~Agram,$^{19}$                                                           
S.H.~Ahn,$^{31}$                                                              
M.~Ahsan,$^{60}$
% ...
\\                                                                            
\vskip 0.30cm                                                                 
\centerline{(D\O\ Collaboration)}                                             
\vskip 0.30cm                                                                 
}                                                                             
\affiliation{                                                                 
\centerline{$^{1}$Universidad de Buenos Aires, Buenos Aires, Argentina}       
\centerline{$^{2}$LAFEX, Centro Brasileiro de Pesquisas F{\'\i}sicas,         
                  Rio de Janeiro, Brazil}                                     
\centerline{$^{3}$Universidade do Estado do Rio de Janeiro,                   
                  Rio de Janeiro, Brazil}                                     
\centerline{$^{4}$Instituto de F\'{\i}sica Te\'orica, Universidade            
                  Estadual Paulista, S\~ao Paulo, Brazil}                     
\centerline{$^{5}$University of Alberta, Edmonton, Alberta, Canada,           
                  Simon Fraser University, Burnaby, British Columbia, Canada,}
\centerline{York University, Toronto, Ontario, Canada, and                    
                  McGill University, Montreal, Quebec, Canada}                
\centerline{$^{6}$Institute of High Energy Physics, Beijing,                  
                  People's Republic of China}                                 
\centerline{$^{7}$University of Science and Technology of China, Hefei,       
                  People's Republic of China}                                 
\centerline{$^{8}$Universidad de los Andes, Bogot\'{a}, Colombia}             
\centerline{$^{9}$Center for Particle Physics, Charles University,            
                  Prague, Czech Republic}                                     
\centerline{$^{10}$Czech Technical University, Prague, Czech Republic}        
\centerline{$^{11}$Center for Particle Physics, Institute of Physics,         
                   Academy of Sciences of the Czech Republic,                 
                   Prague, Czech Republic}                                    
\centerline{$^{12}$Universidad San Francisco de Quito, Quito, Ecuador}        
\centerline{$^{13}$Laboratoire de Physique Corpusculaire, IN2P3-CNRS,         
                   Universit\'e Blaise Pascal, Clermont-Ferrand, France}      
\centerline{$^{14}$Laboratoire de Physique Subatomique et de Cosmologie,      
                   IN2P3-CNRS, Universite de Grenoble 1, Grenoble, France}    
\centerline{$^{15}$CPPM, IN2P3-CNRS, Universit\'e de la M\'editerran\'ee,     
                   Marseille, France}                                         
\centerline{$^{16}$IN2P3-CNRS, Laboratoire de l'Acc\'el\'erateur              
                   Lin\'eaire, Orsay, France}                                 
\centerline{$^{17}$LPNHE, IN2P3-CNRS, Universit\'es Paris VI and VII,         
                   Paris, France}                                             
\centerline{$^{18}$DAPNIA/Service de Physique des Particules, CEA, Saclay,    
                   France}                                                    
\centerline{$^{19}$IPHC, IN2P3-CNRS, Universit\'e Louis Pasteur, Strasbourg,  
                    France, and Universit\'e de Haute Alsace,                 
                    Mulhouse, France}                                         
\centerline{$^{20}$Institut de Physique Nucl\'eaire de Lyon, IN2P3-CNRS,      
                   Universit\'e Claude Bernard, Villeurbanne, France}         
\centerline{$^{21}$III. Physikalisches Institut A, RWTH Aachen,               
                   Aachen, Germany}                                           
\centerline{$^{22}$Physikalisches Institut, Universit{\"a}t Bonn,             
                   Bonn, Germany}                                             
\centerline{$^{23}$Physikalisches Institut, Universit{\"a}t Freiburg,         
                   Freiburg, Germany}                                         
\centerline{$^{24}$Institut f{\"u}r Physik, Universit{\"a}t Mainz,            
                   Mainz, Germany}                                            
\centerline{$^{25}$Ludwig-Maximilians-Universit{\"a}t M{\"u}nchen,            
                   M{\"u}nchen, Germany}                                      
\centerline{$^{26}$Fachbereich Physik, University of Wuppertal,               
                   Wuppertal, Germany}                                        
\centerline{$^{27}$Panjab University, Chandigarh, India}                      
\centerline{$^{28}$Delhi University, Delhi, India}                            
\centerline{$^{29}$Tata Institute of Fundamental Research, Mumbai, India}     
\centerline{$^{30}$University College Dublin, Dublin, Ireland}                
\centerline{$^{31}$Korea Detector Laboratory, Korea University,               
                   Seoul, Korea}                                              
\centerline{$^{32}$SungKyunKwan University, Suwon, Korea}                     
\centerline{$^{33}$CINVESTAV, Mexico City, Mexico}                            
\centerline{$^{34}$FOM-Institute NIKHEF and University of                     
                   Amsterdam/NIKHEF, Amsterdam, The Netherlands}              
\centerline{$^{35}$Radboud University Nijmegen/NIKHEF, Nijmegen, The          
                  Netherlands}                                                
\centerline{$^{36}$Joint Institute for Nuclear Research, Dubna, Russia}       
\centerline{$^{37}$Institute for Theoretical and Experimental Physics,        
                   Moscow, Russia}                                            
\centerline{$^{38}$Moscow State University, Moscow, Russia}                   
\centerline{$^{39}$Institute for High Energy Physics, Protvino, Russia}       
\centerline{$^{40}$Petersburg Nuclear Physics Institute,                      
                   St. Petersburg, Russia}                                    
\centerline{$^{41}$Lund University, Lund, Sweden, Royal Institute of          
                   Technology and Stockholm University, Stockholm,            
                   Sweden, and}                                               
\centerline{Uppsala University, Uppsala, Sweden}                              
\centerline{$^{42}$Physik Institut der Universit{\"a}t Z{\"u}rich,            
                   Z{\"u}rich, Switzerland}                                   
\centerline{$^{43}$Lancaster University, Lancaster, United Kingdom}           
\centerline{$^{44}$Imperial College, London, United Kingdom}                  
\centerline{$^{45}$University of Manchester, Manchester, United Kingdom}      
\centerline{$^{46}$University of Arizona, Tucson, Arizona 85721, USA}         
\centerline{$^{47}$Lawrence Berkeley National Laboratory and University of    
                   California, Berkeley, California 94720, USA}               
\centerline{$^{48}$California State University, Fresno, California 93740, USA}
\centerline{$^{49}$University of California, Riverside, California 92521, USA}
\centerline{$^{50}$Florida State University, Tallahassee, Florida 32306, USA} 
\centerline{$^{51}$Fermi National Accelerator Laboratory,                     
            Batavia, Illinois 60510, USA}                                     
\centerline{$^{52}$University of Illinois at Chicago,                         
            Chicago, Illinois 60607, USA}                                     
\centerline{$^{53}$Northern Illinois University, DeKalb, Illinois 60115, USA} 
\centerline{$^{54}$Northwestern University, Evanston, Illinois 60208, USA}    
\centerline{$^{55}$Indiana University, Bloomington, Indiana 47405, USA}       
\centerline{$^{56}$University of Notre Dame, Notre Dame, Indiana 46556, USA}  
\centerline{$^{57}$Purdue University Calumet, Hammond, Indiana 46323, USA}    
\centerline{$^{58}$Iowa State University, Ames, Iowa 50011, USA}              
\centerline{$^{59}$University of Kansas, Lawrence, Kansas 66045, USA}         
\centerline{$^{60}$Kansas State University, Manhattan, Kansas 66506, USA}     
\centerline{$^{61}$Louisiana Tech University, Ruston, Louisiana 71272, USA}   
\centerline{$^{62}$University of Maryland, College Park, Maryland 20742, USA} 
\centerline{$^{63}$Boston University, Boston, Massachusetts 02215, USA}       
\centerline{$^{64}$Northeastern University, Boston, Massachusetts 02115, USA} 
\centerline{$^{65}$University of Michigan, Ann Arbor, Michigan 48109, USA}    
\centerline{$^{66}$Michigan State University,                                 
            East Lansing, Michigan 48824, USA}                                
\centerline{$^{67}$University of Mississippi,                                 
            University, Mississippi 38677, USA}                               
\centerline{$^{68}$University of Nebraska, Lincoln, Nebraska 68588, USA}      
\centerline{$^{69}$Princeton University, Princeton, New Jersey 08544, USA}    
\centerline{$^{70}$State University of New York, Buffalo, New York 14260, USA}
\centerline{$^{71}$Columbia University, New York, New York 10027, USA}        
\centerline{$^{72}$University of Rochester, Rochester, New York 14627, USA}   
\centerline{$^{73}$State University of New York,                              
            Stony Brook, New York 11794, USA}                                 
\centerline{$^{74}$Brookhaven National Laboratory, Upton, New York 11973, USA}
\centerline{$^{75}$Langston University, Langston, Oklahoma 73050, USA}        
\centerline{$^{76}$University of Oklahoma, Norman, Oklahoma 73019, USA}       
\centerline{$^{77}$Oklahoma State University, Stillwater, Oklahoma 74078, USA}
\centerline{$^{78}$Brown University, Providence, Rhode Island 02912, USA}     
\centerline{$^{79}$University of Texas, Arlington, Texas 76019, USA}          
\centerline{$^{80}$Southern Methodist University, Dallas, Texas 75275, USA}   
\centerline{$^{81}$Rice University, Houston, Texas 77005, USA}                
\centerline{$^{82}$University of Virginia, Charlottesville,                   
            Virginia 22901, USA}                                              
\centerline{$^{83}$University of Washington, Seattle, Washington 98195, USA}  
}                                                                             
\date{June 07, 2006}

\begin{abstract}
A search for leptoquarks has been performed in 310\,\invpb\ of data from
\ppb\ collisions at a center-of-mass energy of 1.96\,TeV, collected by the 
D0 detector at the Fermilab Tevatron Collider. The topology analyzed 
consists of acoplanar jets with missing transverse energy. 
The data show good agreement with standard
model expectations, and a lower mass limit of 136\,\Gev\ has been set at 
the 95\% C.L. for a scalar leptoquark decaying exclusively 
into a quark and a neutrino.
\end{abstract}

\pacs{14.80.-j, 13.85.Rm}
\maketitle 


Many extensions of the standard model (SM) that attempt to explain the 
apparent symmetry between quarks and leptons predict the 
existence of leptoquarks (LQ)\,\cite{LQTh}. 
These new particles are scalar or vector bosons that carry the quantum 
numbers of a quark-lepton system. They are expected to decay into a quark 
and a charged lepton with a branching fraction $\beta$, or into a quark and 
a neutrino with a branching fraction $(1-\beta)$. 
At \ppb\ colliders, leptoquarks can be pair produced, if sufficiently light,
primarily by $q\bar q$ annihilation and gluon-gluon fusion, with a production
cross section independent of the unknown leptoquark-quark-lepton coupling.
For $\beta = 0$, the resulting final state consists of a pair of 
acoplanar quark jets with missing transverse energy, \met, carried away by the 
two neutrinos. 

In this Letter, a search for leptoquarks that decay into a quark 
and a neutrino, using data collected at a center-of-mass energy of 1.96\,TeV 
with the D0 detector during Run~II of the Fermilab Tevatron Collider, is 
reported. The production cross section for vector leptoquark pairs is larger 
than that for scalar leptoquarks, but it is model
dependent. The interpretation of the results is therefore presented in terms of
scalar leptoquark masses. 
The most constraining 95\% C.L. lower mass limit for $\beta=0$, previous to 
this search, 
was 117\,\Gev, obtained by the CDF Collaboration with 191\,\invpb~of Run II 
data\,\cite{CDF}.

A detailed description of the D0 detector can be found in 
Ref.~\cite{detector}. The central tracking system consists of a
silicon microstrip tracker and a fiber tracker,
both located within a 2~T superconducting solenoidal magnet. 
A liquid-argon and uranium calorimeter covers pseudorapidities 
up to $|\eta|$ $\approx 4.2$, where 
$\eta=-\ln \left[ \tan \left( \theta/2 \right) \right]$ and $\theta$ 
is the polar angle with respect to the proton beam direction. 
The calorimeter consists of three sections housed in separate cryostats: 
the central one covers $|\eta|$ $\lsim 1.1$, and the two end 
sections extend the coverage to larger $\vert\eta\vert$. The calorimeter 
is segmented in depth, with four electromagnetic layers followed by up to 
five hadronic layers. It is  also segmented in projective towers of size 
$0.1\times 0.1$ in $\eta$--$\phi$ space, where $\phi$ is the azimuthal 
angle in radians. Calorimeter cells are formed by the intersections of towers 
and layers. Additional energy sampling is provided by scintillating tiles 
between cryostats.
An outer muon system, covering $|\eta|<2$, consists of a layer of tracking 
detectors and scintillation trigger counters in front of 1.8~T iron toroids, 
followed by two similar layers beyond the toroids. 

For this search, data collected with a \mbox{jets + $\met$} trigger have been 
analyzed.
At the first level, this trigger selects events in which at least three 
calorimeter trigger
towers of size $\Delta\phi \times \Delta\eta\,=\,0.2 \times 0.2$ 
record a transverse energy in excess of 5\,GeV. 
At the second and third trigger
levels, requirements are placed on $\mht$, the vector sum of the jet transverse
momenta ($\mht = \vert\sum_{\text{jets}} \overrightarrow p_T\vert$). 
Coarse jets are reconstructed from trigger towers at the second level, 
while the full detector information is used at the third level. 
The $\mht$ thresholds are 20 and 30\,GeV at the second and third levels, 
respectively.
The trigger efficiency is larger than 98\% for events fulfilling the selection
criteria of this analysis. 
Data quality requirements on the performance of each detector subsystem 
yielded an available integrated luminosity of 310\,\invpb.

The offline analysis utilized jets reconstructed with the iterative midpoint 
cone algorithm\,\cite{jetalgo} with a cone size of 0.5.
The jet energy calibration was derived from the transverse momentum balance in 
photon+jet events. 
Only jets with $p_T > 15$\,GeV that passed general quality criteria, 
based on the jet longitudinal profile in the calorimeter, were selected 
for this analysis.
The mis\-sing
transverse energy was calculated from all calorimeter cells, corrected for
the energy calibration of reconstructed jets and for the momentum of
reconstructed muons. 

The sample of approximately 14 million events collected with the jets + $\met$ 
trigger was reduced by requiring the following preselection 
criteria to be satisfied: at least two jets; $\mht>40$\,GeV; 
$\met>40$\,GeV, where, in contrast 
to \mht,  information from energy not belonging to 
reconstructed jets is taken into account; 
and $\Delta\Phi < 165^\circ$,
where $\Delta\Phi$ is the acoplanarity of the two leading jets, i.e., 
the two jets with the largest transverse momenta,
defined as the difference between their azimuthal angles. 
To ensure that the selected events were well contained in the detector, the 
position of the interaction vertex along the beam direction was required to be 
within 60\,cm of the detector center.

Events in which the presence of obvious calorimeter noise could be detected 
were rejected. The inefficiency associated 
with this procedure was measured using events collected at random beam 
crossings (zero-bias events), and events collected with an unbiased 
trigger and containing exactly two jets back-to-back in azimuth. 
At this stage, 306,\,937 events survived. 

Signal efficiencies and SM backgrounds have been evaluated 
using a full {\sc geant}\,\cite{GEANT} based simulation of events, 
with a 
Poisson average of 0.8 minimum-bias events superimposed, corresponding to the
luminosity profile in the data sample analyzed. These 
simulated events were
reconstructed in the same way as the data. The jet 
energies further received calibration corrections and an additional 
smearing to 
take into account residual differences between data and 
simulation, as determined with photon+jet events. The 
instrumental background due to jet energy 
mismeasurements in QCD multijet production 
was estimated directly from the data.

The SM processes expected to yield the largest background contributions are 
vector boson production in association with jets, among which $Z\to\nu\nu$ is
irreducible. Vector boson pair production and top quark production have also 
been considered. All of these processes
were generated with {\sc alpgen 1.3}\,\cite{ALPGEN}, interfaced with 
{\sc pythia 6.202}\,\cite{PYTHIA} for the simulation of initial and final state
radiation and for jet hadronization. 
The parton distribution functions (PDFs) used were CTEQ5L\,\cite{PDF5}.
The next-to-leading order (NLO) cross sections for vector boson production in
association with jets were calculated with {\sc mcfm 3.4.4}\,\cite{MCFM} and
the CTEQ5M PDFs.

The production of scalar leptoquarks via the processes
\qqb\ or $gg$ $\to$ ${\mathrm{LQ}}\overline{\mathrm{LQ}}$
was simulated with {\sc pythia} and the CTEQ5L PDFs. 
The chosen leptoquark masses ranged from 80 to 140\,\Gev, in steps of 5\,\Gev.
For each mass, 10,000 events were generated.
The NLO leptoquark pair production cross sections were calculated using a 
program 
based on Ref.\,\cite{Kraemer}, with CTEQ6.1M PDFs\,\cite{PDF6}. For the 
mass range considered, they vary from 52.4 to 2.38\,pb. These nominal values 
were obtained for a renormalization and factorization scale equal to the
leptoquark mass.

The selection criteria for this analysis are listed in 
Table~\ref{lqeffcuts}, together with the numbers of events surviving at 
each step and with the cumulative efficiency for a leptoquark mass 
of 140\,\Gev.
The jet kinematic cuts {\bf C1} to {\bf C4} reject a large fraction of the 
SM and instrumental backgrounds. 
They take advantage of the central signal production 
and decay by requiring that \etadet\ be smaller than 1.5
for the two leading jets, 
where $\eta_\mathrm{det}$ is the pseudorapidity measured from the detector 
center. Cut {\bf C5}, where EMF is the fraction of jet energy contained in 
the electromagnetic section of the calorimeter, rejects jets likely due
to photons or electrons. 

\begin{table}
\caption{\label{lqeffcuts} Numbers of data events selected and signal 
cumulative efficiencies for $m_{LQ} = 140$\,\Gev\ at various stages of the 
analysis. The leading and subleading jets are denoted jet-1 and jet-2.}
\begin{ruledtabular}
\begin{tabular}{lrr}
Cut applied & Events left & Signal eff. (\%) \\
\hline
Initial cuts & 306,937 & 58.8 \\
{\bf C1:} jet-1 $\pt > 60$\,\Gev & 206,116 & 48.7 \\
{\bf C2:} jet-1 $\etadet < 1.5$ & 160,323 & 46.8 \\
{\bf C3:} jet-2 $\pt > 50$\,\Gev\ & 48,979 & 24.8 \\
{\bf C4:} jet-2 $\etadet < 1.5$ & 42,028 & 22.7 \\
{\bf C5:} jet-1 jet-2 $\mathrm{EMF} < 0.95$ & 40,821 & 22.3 \\ 
{\bf C6:} jet-1 jet-2 $\mathrm{CPF} > 0.05$ & 34,746 & 22.2 \\
{\bf C7:} exactly two jets & 5,213 & 15.3 \\
{\bf C8:} $\met > 70$\,GeV & 492 & 11.8 \\
{\bf C9:} isolated electron veto& 465 & 11.7 \\
{\bf C10:} isolated muon veto& 399 & 11.6 \\
{\bf C11:} isolated track veto& 287 & 10.0 \\
{\bf C12:} $\dphimax-\dphimin < 120^\circ$ & 180 & 9.4 \\
{\bf C13:} $\dphimax+\dphimin < 280^\circ$ & 124 & 8.4 \\
{\bf C14:} $\met > 80$\,GeV & 86 & 7.0 \\
\end{tabular}
\end{ruledtabular}
\end{table}

In cut {\bf C6}, 
the total transverse energy of the charged particles emanating from the 
interaction vertex and associated with a jet, as measured in the tracking 
system, 
is compared to the jet transverse energy recorded in the calorimeter. 
The charged particle fraction CPF, i.e. the ratio of these two quantities, 
is expected to be close to zero either if a wrong interaction vertex 
was selected, 
in which case it is unlikely that the charged tracks truly associated with the 
jet will come from the selected vertex, or if the jet is a fake one,
e.g. due to calorimeter noise, 
in which case there should be no real charged tracks associated with it. 
The efficiency of this jet confirmation procedure was determined using 
events containing two jets back-to-back in azimuth.

Cut {\bf C7} was applied to suppress further the instrumental background, 
which is enriched in multijet events by the acoplanarity requirement. 
The efficiency of such a jet multiplicity cut is sensitive to the modeling of 
initial and final state radiation (ISR/FSR). To verify the simulation of these 
effects, $(Z\to ee)+\geq 2$-jet events were selected in the data, and compared 
to a simulation by {\sc alpgen} for the production of $(Z\to ee)$+2-jets, with 
ISR/FSR jets added by {\sc pythia}. The two leading jets were required to 
fulfil criteria similar to those used in the analysis, and the numbers of 
events with additional jets were compared between data and simulation, as well 
as the \pt\ spectra of those jets. The small deficit observed in the 
simulation, located mostly at $\pt<20$\,GeV, was used to correct the signal and
background simulations, and the statistical power of this test was taken as a 
systematic uncertainty.

After cut {\bf C8}, the level of the 
instrumental background is largely reduced and is 
similar to the level of the SM backgrounds. 
The final \met\ cut value (cut {\bf C14}) was optimized as 
explained below. 

Cuts {\bf C9}, {\bf C10} 
and {\bf C11}, reject a large fraction of the events originating from 
$W/Z$+jet processes. In cut {\bf C9}, an electron with $p_T>10$\,GeV is 
declared isolated if the calorimeter energy in a cone of radius 0.4 in 
$\eta$--$\phi$ around the electron direction does not exceed the energy contained 
in the electromagnetic layers inside a cone of radius 0.2 by more than 15\%. 
In cut {\bf C10}, a muon with $p_T>10$\,GeV is declared isolated if the 
calorimeter energy in a hollow cone with inner and outer radii 0.1 and 0.4
around the muon direction is smaller than 2.5\,GeV, and if the sum of the
transverse energies of charged tracks, other than the muon, in a cone of 
radius 0.5 is smaller than 2.5\,GeV. In cut {\bf C11}, a charged track with 
$p_T>5$\,GeV is declared isolated if no charged track with $p_T>0.5$\,GeV is
found within a hollow cone of radii 0.1 and 0.4 around the track considered. 
This cut was specifically 
designed to reject hadronic decays of $\tau$-leptons; the use of a hollow, 
rather than full cone renders it efficient also in case of decays
into three charged particles.

\begin{figure}
\includegraphics[width=8.5cm]{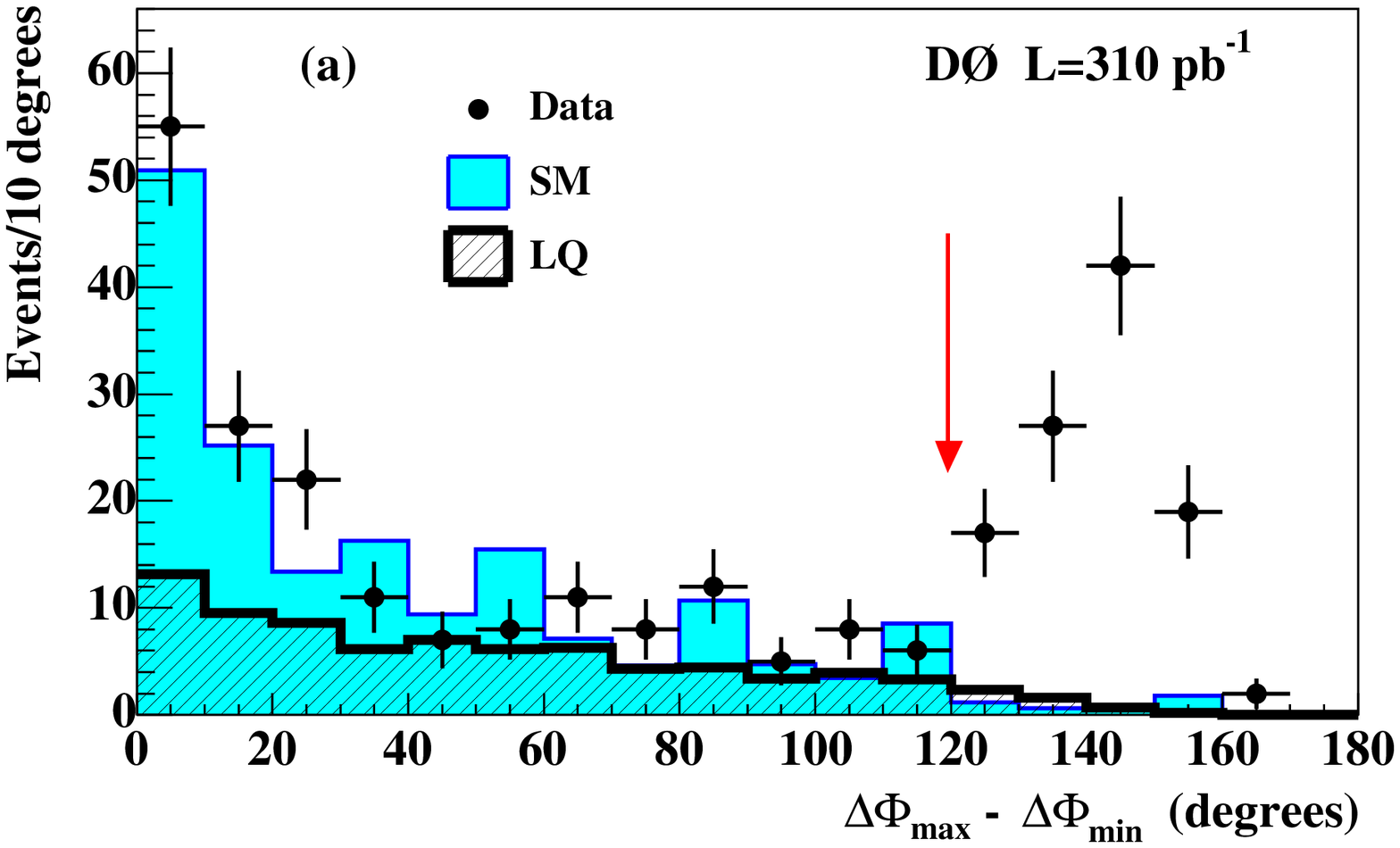} \\
\includegraphics[width=8.5cm]{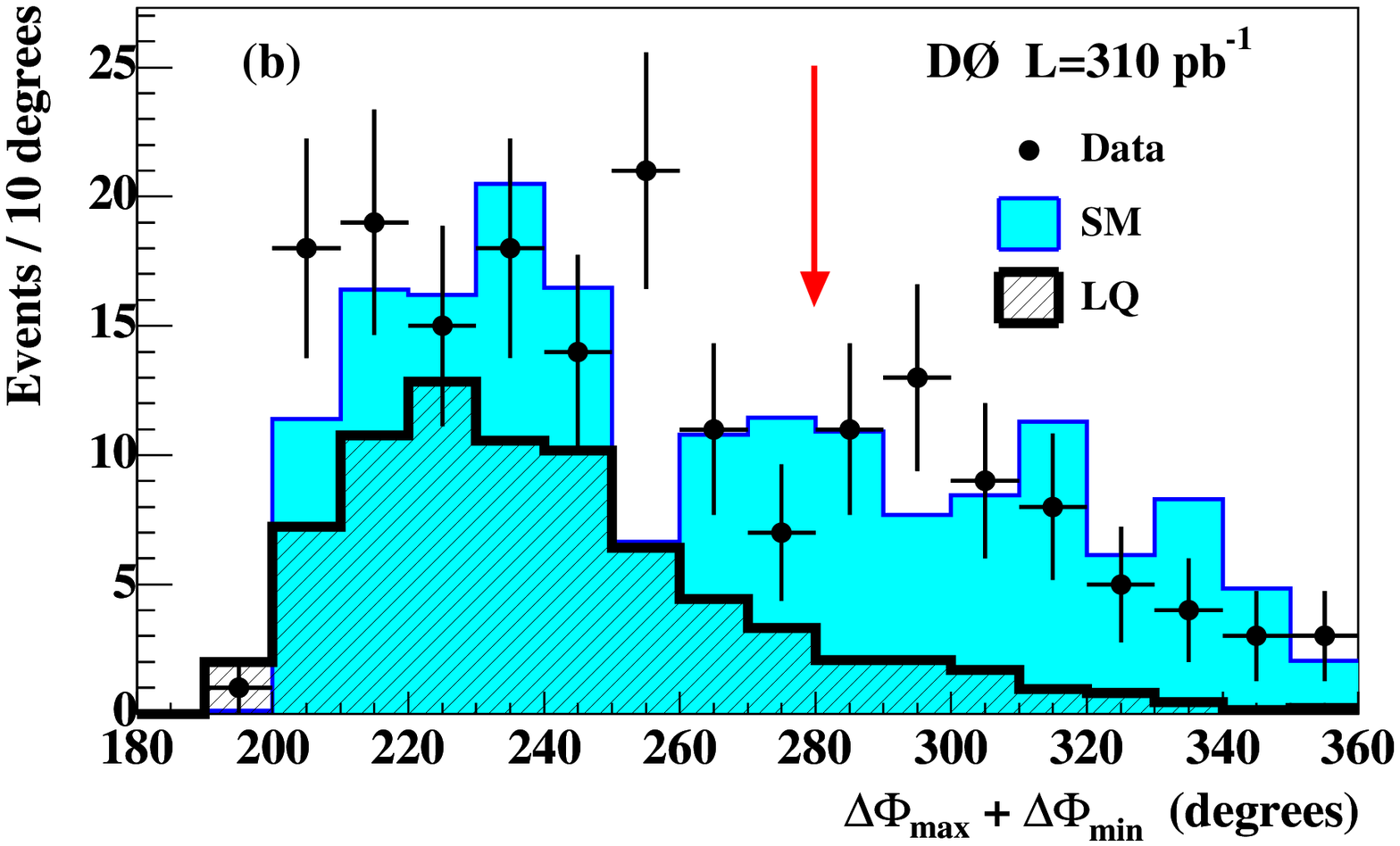}
\caption{\label{angles} Distributions of $\dphimax-\dphimin$ (a) and of 
$\dphimax+\dphimin$ (b) for data (points with error bars), for SM
backgrounds (shaded histograms), and for a 140\,\Gev\ LQ signal 
(hatched histograms). 
In the $\dphimax-\dphimin$ distribution, cuts {\bf C1} to {\bf C11} are 
applied. The excess in data beyond $120^\circ$
is attributed to the non-simulated instrumental background. 
In the $\dphimax+\dphimin$ distribution, the cut 
$\dphimax-\dphimin < 120^\circ$ ({\bf C12}) has been applied in addition. 
The locations of cuts 
{\bf C12} and {\bf C13} are indicated by arrows in (a) and (b), respectively.}
\end{figure}

The angular correlations between the jet and \met\ directions are used to
suppress both the instrumental and SM backgrounds. To this end, the
minimum $\Delta\Phi_\mathrm{min}(\met,\mathrm{any~jet})$ and maximum 
$\Delta\Phi_\mathrm{max}(\met,\mathrm{any~jet})$  of the azimuthal angle 
differences between the\,~\met\ direction and the direction of any of 
the two jets are combined as shown in Fig.\,\ref{angles}. 
It can be seen in Fig.\,\ref{angles}a that cut {\bf C12} rejects most of the 
remaining instrumental background, which is responsible for the excess beyond
$120^\circ$. Cut {\bf C13}, which suppresses SM 
backgrounds at the expense of a moderate reduction of the signal efficiency,  
was optimized as explained below. The variable $\dphimax+\dphimin$ is the one
which discriminates best the signal and the irreducible background from 
$(Z\to\nu\nu)$+2-jets. Its effect is demonstrated in Fig.\,\ref{angles}b.

\begin{figure}
\includegraphics[width=8.5cm]{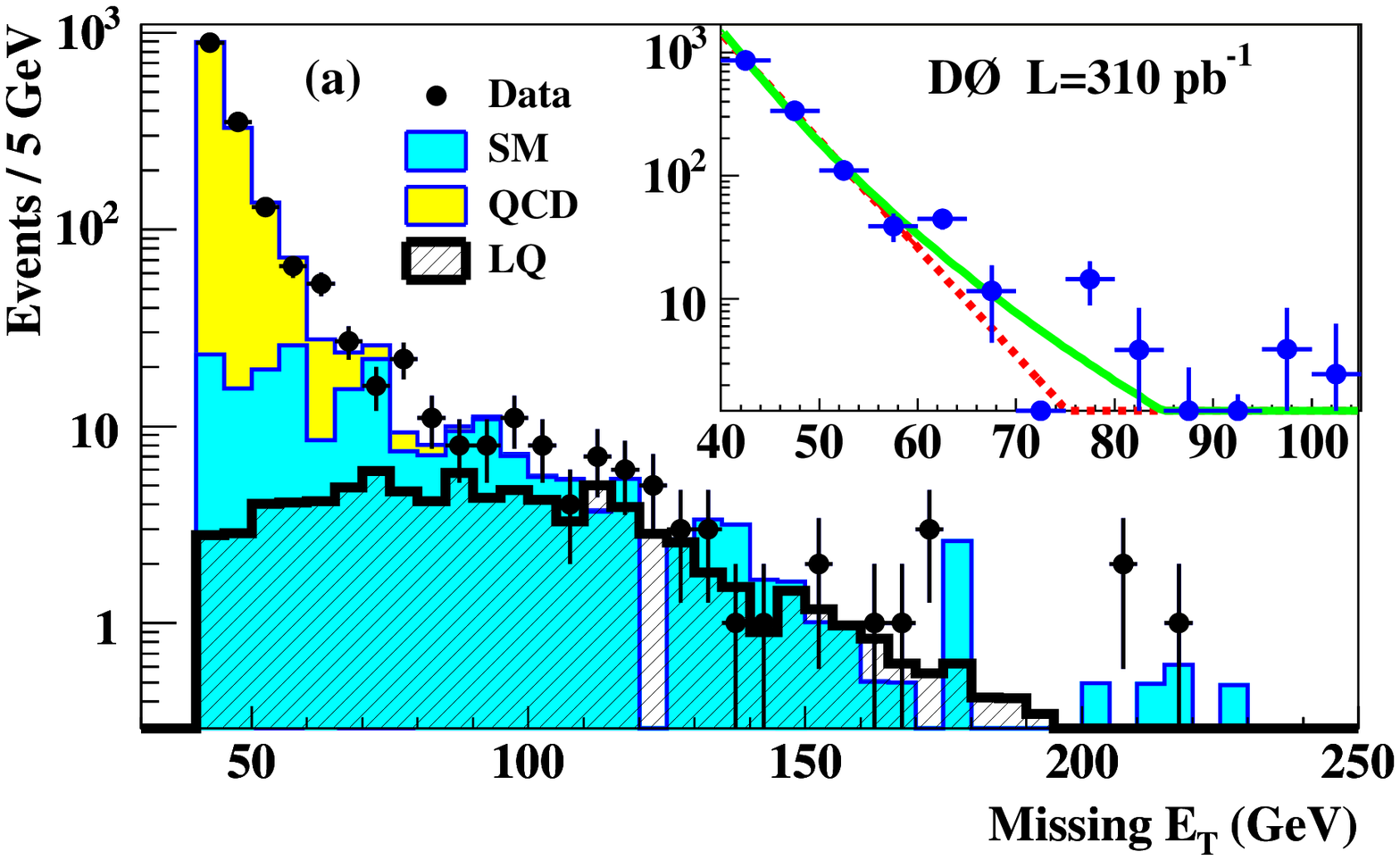} \\
\includegraphics[width=8.5cm]{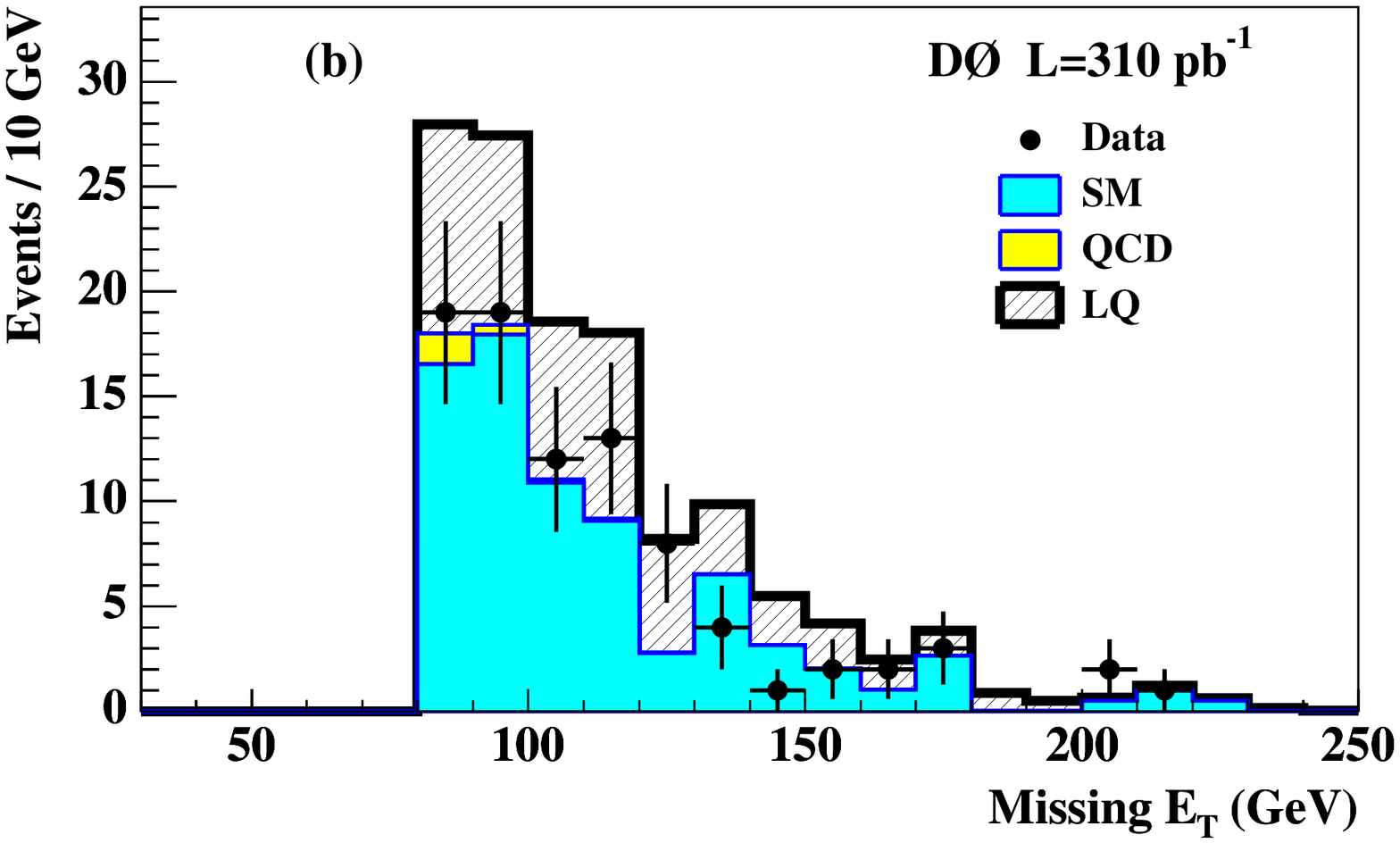}
\caption{\label{lqmet} Distributions of \met\ 
for data (points with error bars), for SM backgrounds (heavy-shaded 
histograms), for the instrumental background 
(labeled QCD, light-shaded  histograms), and 
for a 140\,\Gev\ LQ signal. 
In (a), all cuts except {\bf C8} and {\bf C14} are applied, the LQ signal is 
shown 
as a hatched histogram, and the insert shows how the instrumental background
contribution 
is estimated from power law (solid curve) and exponential (dashed curve) fits. 
The \met\ distribution in (b) is after all cuts, with the same 
shading code but with the signal contribution now displayed on top of all 
backgrounds.}
\end{figure}

Finally, the $\met$ and $\dphimax+\dphimin$ cuts were optimized for a 140\,GeV
LQ mass so as to minimize the cross section expected to be excluded in the 
absence of signal. 
Cut {\bf C8} was removed, and \met\ cut values ranging from 60 
to 90\,GeV were probed in 10\,GeV steps. The cut on $\dphimax+\dphimin$ was 
varied 
between $260^\circ$ and $300^\circ$ in steps of $10^\circ$. For each set of 
cuts, the instrumental background was estimated as explained below. 
The systematic 
uncertainties discussed further down were taken into account in the 
calculation of the expected limits. The optimal set of cuts reported
as {\bf C13} and {\bf C14} in Table~\ref{lqeffcuts} selects 86 data events.

The instrumental background was estimated from exponential and power law fits 
to the \met\ distribution (insert of Fig.\,\ref{lqmet}a) in the range 
$[40,60]$\,GeV, where the signal contribution is negligible,
after subtraction of the SM expectation. Both fits were 
extrapolated beyond the \met\ cut value, and the average of the two results was
taken as the instrumental background estimate, with a systematic uncertainty 
accounting for the difference between the two fit results. The final \met\
distribution is shown in Fig.\,\ref{lqmet}b.
The values of the SM and instrumental backgrounds are 
given in Table~\ref{lqsmbg}. 
The largest background sources are, as expected, 
$(Z\to\nu\nu)$+2-jets and 
$(W\to\ell\nu)$+jets \mbox{($\ell=e$, $\mu$, $\tau$).}

\begin{table}
\renewcommand{\arraystretch}{1.4}
\caption{\label{lqsmbg} Numbers of events expected from standard model, 
instrumental and total backgrounds; 
number of data events selected; and number of signal events expected for
$m_{LQ}=140$\,\Gev, assuming the nominal production cross section. For the 
total SM and total backgrounds, as well as for the signal, the first 
uncertainties are statistical and the second systematic. 
The uncertainties on the individual SM
backgrounds are statistical. The uncertainty on the instrumental background
is mostly systematic from the difference between the power law and exponential 
fits.}
\begin{ruledtabular}
\begin{tabular}{lc}
$(Z\to\nu\nu)$+2-jets & $34.6 \pm 4.3$ \\
$(W\to\ell\nu)$+jets & $35.0 ^{+9.1}_{-8.7}$ \\
$(Z\to\ell\ell)$+jets & $0.3 ^{+0.4}_{-0.2}$ \\
$t\bar t$ & $1.9 \pm 0.1$ \\
$WW$, $WZ$, $ZZ$ & $1.2 \pm 0.2$ \\
\hline 
Total SM background & 72.9 \mbox{$^{+10.1}_{-9.7}$} \mbox{$^{+10.6}_{-12.1}$}  \\
Instrumental background & $ 2.3 \pm 1.2 $ \\
\hline
Total background & 75.2 \mbox{$^{+10.1}_{-9.7}$} \mbox{$^{+10.7}_{-12.2}$} \\  
\hline
Data events selected & 86 \\
\hline
Signal ($m_{LQ}=140$\,\Gev) & $ 51.8 \pm 1.8 ^{+5.6}_{-4.6} $ \\
\end{tabular}
\end{ruledtabular}
\end{table}

The signal efficiencies at various stages of the analysis are 
given in Table~\ref{lqeffcuts} for $m_{LQ} = 140$\,\Gev. The efficiency
decreases together with the leptoquark mass, reaching 1.6\% at 100\,\Gev. 
The number of signal events expected for a 
leptoquark mass of 140\,\Gev\ is indicated in Table~\ref{lqsmbg}.

The following sources of systematic uncertainty are fully correlated between 
SM background and signal expectations:
the relative jet energy calibration between data and simulation: 
$^{+4}_{-8}$\% for the SM background and $^{+6}_{-4}$\% for the signal;
the relative jet energy resolution between data and 
simulation: $^{+2}_{-4}$\% for the SM background and negligible
for the signal;
the efficiency of the jet multiplicity cut: $\pm 3$\%,
after corrections of $-3$\% for the SM background and $-2$\% for the signal; 
the trigger efficiency: $\pm 2$\% after all selection cuts;
and the integrated luminosity of the analysis sample: $\pm 6.5$\%.

In addition to the $^{+14}_{-13}$\% statistical uncertainty of the simulation, 
the normalization of the SM background expectation is affected by a $\pm 12$\% 
uncertainty, as inferred from a comparison of data and simulated 
$(Z\to ee)$+2-jet events selected with the same criteria for the jets as in the
analysis sample.
The uncertainty of $\pm 1.2$ events on the instrumental background was 
estimated
from power law and exponential fits to the \met\ distribution, as explained 
previously.
As a check, the same procedure was applied to the events with
$\dphimax-\dphimin > 120^\circ$, which are dominated by the instrumental 
background contribution. 
This showed that the high\,~\met\ tail is somewhat underestimated, possibly by
as much as nine events, which
leads to conservative results in terms of limit setting. Finally,
the uncertainty on the signal efficiency due to the PDF choice was 
determined to be $^{+6}_{-4}$\%, 
using the twenty-eigenvector basis of the CTEQ6.1M PDF set\,\cite{PDF6}.

\begin{figure}
\includegraphics[width=8.5cm]{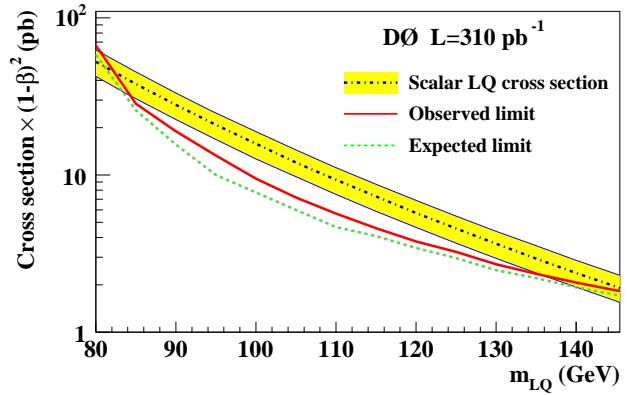}
\caption{\label{exclu} Observed (solid curve) and expected (dashed curve) 
95\% C.L. upper limits on the cross section times $(1-\beta)^2$ as functions 
of the leptoquark mass. 
The nominal cross section for scalar-leptoquark pair production is also 
shown for $\beta=0$ (dash-dotted curve), 
with the shaded band indicating the uncertainty due to the choices of PDFs and
of renormalization and factorization scale.} 
\end{figure}

As can be seen in Table~\ref{lqsmbg} and Fig.\,\ref{lqmet}b, 
no significant excess of events is 
observed in the data above the background expectation. Therefore,
given the number of selected events, 
the SM and instrumental background expectations,
the integrated luminosity of 310\,\invpb, the signal selection efficiency 
as a function of the leptoquark mass,
and the statistical and systematic uncertainties discussed above, a 95\% C.L. 
upper limit on the cross section times $(1-\beta)^2$ has been determined 
as shown in Fig.\,\ref{exclu},
using the modified frequentist $CL_s$ approach\,\cite{CLs}.
The expected limit in the absence of signal is also indicated. 

The nominal theoretical cross section for the pair production of scalar 
leptoquarks is also shown in Fig.\,\ref{exclu}. It was obtained based on 
Ref.\,\cite{Kraemer} with CTEQ6.1M PDFs and for
a renormalization and factorization scale $\mu_{rf}$ equal to the leptoquark 
mass. The uncertainty associated with the PDF choice was estimated using the 
full set of CTEQ6.1M eigenvectors and combined 
quadratically with the variations obtained when $\mu_{rf}$ was modified by a 
factor of two up or down. For a leptoquark
mass of 140\,\Gev, the PDF uncertainty on the theoretical cross section amounts
to $^{+18}_{-13}$\% and the scale variation results in a change of 
$^{+11}_{-13}$\%, the quadratic sum being $^{+21}_{-19}$\%. 
Reducing the nominal cross section by this theoretical uncertainty, shown as 
the shaded band in Fig.\,\ref{exclu}, a lower mass limit of 
136\,\Gev\ is derived at the 95\% C.L. Masses smaller than 85\,\Gev, to 
which this analysis is not sensitive, have been excluded 
previously\,\cite{CDF,LQRev}. 
The cross section limit obtained here was combined with the results of the
published D0 search for first-generation scalar leptoquarks in the $eeqq$ and 
$e\nu qq$ final states\,\cite{DLQ1}, and the lower mass limit of 136\,GeV 
was seen to hold independent of $\beta$. 

In summary,
a search for acoplanar jet final states in \ppb\ collisions at 1.96\,TeV, 
performed using a data sample of 310\,\invpb\ collected by the D0 detector, 
revealed no deviation from the standard model expectation. 
For a single-generation  scalar leptoquark, a 
lower mass limit of 136\,\Gev\ has been obtained for $\beta=0$. 
While a tighter limit is 
available for third-generation leptoquarks\,\cite{CDF3}, due to the increased 
signal purity achieved with heavy flavor tagging, this is the most stringent 
limit to date for first- and second-generation scalar leptoquarks decaying 
exclusively into a quark and a neutrino.

%
We thank the staffs at Fermilab and collaborating institutions, 
and acknowledge support from the 
DOE and NSF (USA);
CEA and CNRS/IN2P3 (France);
FASI, Rosatom and RFBR (Russia);
CAPES, CNPq, FAPERJ, FAPESP and FUNDUNESP (Brazil);
DAE and DST (India);
Colciencias (Colombia);
CONACyT (Mexico);
KRF and KOSEF (Korea);
CONICET and UBACyT (Argentina);
FOM (The Netherlands);
PPARC (United Kingdom);
MSMT (Czech Republic);
CRC Program, CFI, NSERC and WestGrid Project (Canada);
BMBF and DFG (Germany);
SFI (Ireland);
The Swedish Research Council (Sweden);
Research Corporation;
Alexander von Humboldt Foundation;
and the Marie Curie Program.
%

\end{document}